\documentclass[aps,reprint,twocolumn,prl,showpacs,superscriptaddress,floatfix]{revtex4-1}
\usepackage{amssymb,amsmath,amsfonts} 
\usepackage{epsfig,graphicx}

\begin{document}

\title{Approximated integrability of the Dicke model}


 \author{A. Rela\~{n}o}
  \affiliation{Departamento de F\'{\i}sica Aplicada I and GISC,
    Universidad Complutense de Madrid, Av. Complutense s/n, 28040
    Madrid, Spain} \email{armando.relano@fis.ucm.es}
  \author{M. A. Bastarrachea-Magnani} \affiliation{Instituto de Ciencias Nucleares, Universidad Nacional Aut\'onoma de M\'exico,
Apdo. Postal 70-543, M\'exico D. F., C.P. 04510}\email{miguel.bastarrachea@nucleares.unam.mx}
\author{S. Lerma-Hern\'andez}\affiliation{Instituto de Ciencias Nucleares, Universidad Nacional Aut\'onoma de M\'exico,
Apdo. Postal 70-543, M\'exico D. F., C.P. 04510}\affiliation{Facultad  de F\'\i sica, Universidad Veracruzana,
Circuito Aguirre Beltr\'an s/n, Xalapa, Veracruz, M\'exico, C.P. 91000}\email{slerma@uv.mx}

\begin{abstract}
  A very approximate second integral of motion of the Dicke model is
  identified within a broad region above the ground state, and for a
  wide range of values of the external parameters. This second
  integral, obtained from a Born Oppenheimer approximation, classifies
  the whole regular part of the spectrum in bands labelled by its
  corresponding eigenvalues. Results obtained from this approximation
  are compared with exact numerical diagonalization for finite systems
  in the superradiant phase, obtaining a remarkable accord. The region
  of validity of our approach in the parameter space, which includes
  the resonant case, is unveiled. The energy range of validity goes
  from the ground state up to a certain upper energy where chaos sets
  in, and extends far beyond the range of applicability of a simple
  harmonic approximation around the minimal energy configuration.  The
  upper energy validity limit increases for larger values of the
  coupling constant and the ratio between the level splitting and the
  frequency of the field. These results show that the Dicke model
  behaves like a two-degree of freedom integrable model for a wide
  range of energies and values of the external parameters.
\end{abstract}

\maketitle

{\em Introduction.-} Despite it was formulated more than $60$ years
ago, the Dicke model \cite{Dicke:54} has been object of intense
research during the last years. One of the reasons is that it shows a very rich and complex
behavior, in spite of being quite simple. It is characterized by
thermal \cite{Hepp:73,Bastarrachea:16}, quantum (QPT) \cite{Emary:03}
and excited-state quantum (ESQPT) \cite{Perez-Fernandez:11} phase
transitions, giving rise to superradiance. These phase transitions are
also extended to non-equilibrium driven systems
\cite{Bastidas:12}. Besides, the Dicke model shows a transition from
regularity to chaos, observable both in the normal and the superradiant
phases \cite{Bastarrachea:14}.

Theoretical studies on this model are enhanced by recent experimental results. In cavity QED, the superradiant phase transition is forbidden by a no-go theorem \cite{Nataf:10}. However, this kind of 
transition has been observed in several systems \cite{Hartmann:06}, and the Dicke model itself has been simulated by means of a Bose-Einstein condensate in an optical cavity \cite{Baumann:10}. Dynamical non-equilibrium superradiant phase transition has been also observed \cite{Klinder:15}. The same model with just one atom, the Rabi model, has been explored by means of superconducting QED \cite{Niemczyk:10}. Finally, similar techniques have been applied to few-atoms Dicke models \cite{Casanova:10}.

For all these reasons, a precise knowledge of the spectrum of the Dicke model is a theoretical challenge and a necessity. The low-lying spectrum can be approximated by two independent harmonic oscillators through a Holstein-Primakoff approximation in the thermodynamical limit \cite{Emary:03}. However, specially in the superradiant phase, this method only works in a very narrow region around the ground-state, leaving unexplained  the whole non-chaotic or regular energy region that extends far beyond the validity of the quadratic approximation. This regular energy region extends from the ground to an upper energy (above or below the ESQPT critical energy) that depends strongly on the   level splitting and frequency of the field \cite{Chavez:16}. On the other hand, it has been recently shown that the Rabi model can be considered as integrable \cite{Braak:11}, though there is some controversy about this fact \cite{Moroz:13}. A method similar to the one used in \cite{Braak:11} has been recently applied to the Dicke model with two atoms \cite{Duan:15}, but it does not generate a closed formula for the spectrum. This should be expected as the Hamiltonian does not have as many integrals of motion as degrees of freedom, and thus in that sense it is non-integrable. Therefore, a general concise analytic solution of the Dicke model is far to be obtained.

In this letter we derive explicitly  a second integral of motion of the  Dicke model, very approximately holding within its whole non chaotic energy regime and a wide range of values of the external parameters, specially in the superradiant region.
Its range of applicability is shown  analytically and  by means of stringent numerical calculations. 


Our derivation is based on a Born-Oppenheimer approximation (BOA) \cite{Berry:93} where the fast and slow variables are, respectively,  the atomic and bosonic ones. Similar approaches have already been  reported for the Dicke and related models. Regarding the Dicke model, the BOA has been used to study its ground state properties. In Ref.\cite{Liberti:06} the finite size scaling of the entanglement between the components of the system and  other physical obervables is reported, and in Ref.\cite{Chen:07} the BOA is employed to study the finite size dependence of the tunnelling driven ground state energy splitting in the superradiant phase. 
The BOA has, likewise,  been applied to the one atom   Dicke model, the Rabi and its rotated-wave approximated Jaynes-Cummings model. In Ref.\cite{Liberti2:06} the BOA is used to determine the entanglement of the single atom to the bosonic variable, whereas in Ref.\cite{Larson:07} it is used to unveil some previously unnoticed aspects of the rotating wave approximation. A different and complementary  BOA, where the fast variables are the bosonic ones, is reported in \cite{Irish:05}.  

This letter is organized as follows.  After presenting the BOA and the analysis of its regions of applicability in the model parameter space,  we give numerical evidence showing that the whole regular part of the Dicke dynamics can be understood from the adiabatic invariant coming from the BOA. Finally, we show the ability of the BOA to reproduce the regular exact Dicke spectrum, and that, contrary to the quadratic or harmonic approximation, the accuracy of the BOA increases with the number of atoms.    



{\em The Dicke model and its second integral of motion.-} The Dicke model depicts the interaction of $N$ two-level atoms with a single bosonic mode. The atoms can be described by means of collective
pseudospin operators, giving rise to ($\hbar=1$)
\begin{equation}
H = \omega a^{\dagger} a + \omega_0 J_z + \frac{2 \gamma}{\sqrt{2j}} J_x \left( a^{\dagger} + a \right),
\end{equation}
where $\omega$ is the frequency of the bosonic mode, $\omega_0$ the level splitting, $\gamma$ the coupling constant, and $j$ the total pseudospin $j=N/2$. This model can be numerically solved for tens or even few hundreds \cite{Bastarrachea:11} of atoms, depending on $\gamma$. The main difficulty is that the bosonic Hilbert space is infinite and has to be truncated for an exact diagonalization; the larger the coupling constant, the more bosons are required for convergence.

Here we develop a very approximated solution using 
a Born-Oppenheimer approximation, valid when a fast system is coupled to a much slower one. When this condition is fulfilled, the fast motion can be solved for frozen (adiabatically changing) values of the slow coordinates; and then solved for the slow motion, by considering temporal averages of the fast coordinates.  We apply this idea to the Dicke model, considering that the atomic are the fast coordinates, whereas the bosonic are the slow ones.
To implement this approach, we consider a semiclassical approximation for the slow variables, $a =
(1/\sqrt{2}) (q + i p)$, from we obtain
\begin{equation}
H(p,q) =  \frac{\omega}{2} \left( p^2 + q^2 \right) +\omega_0 J_z + \frac{2 \gamma}{\sqrt{j}} q J_x.
\end{equation}
If we consider $q$ and $p$ as fixed parameters, this Hamiltonian can be exactly diagonalized by means of a rotation around the $y$-axis,
\begin{equation}
  \omega_0 J_z + \frac{2 \gamma}{\sqrt{j}} q J_x = \sqrt{\omega_0^2 + \left( \frac{2 \gamma q}{\sqrt{j}} \right)^2} J_{z'}\equiv \omega_P(q)\,J_{z'}.
\label{eq:jz}
\end{equation}
Considering that the critical coupling of the quantum phase transition
is $\gamma_c=\sqrt{\omega \omega_0}/2$, this rotation entails 
\begin{equation}
  H(p,q) = \frac{\omega}{2} \left( p^2 + q^2 \right) + \omega_0\sqrt{1+\frac{\omega}{\omega_0 j} \left(\frac{\gamma}{\gamma_c} \right)^2 q^2} J_{z'}.
\end{equation}
This Hamiltonian describes the Larmor precession, with frequency $\omega_P(q)$, of the pseudospin around an adiabatically changing $q$-dependent axis. By applying the Hamiltonian to the eigenbasis of $J_{z'}$, $J_{z'} \left| j,m' \right> = m' \left| j,m' \right>$, we obtain the eigenvalues
\begin{equation}
E_{m'} (p,q) = \frac{\omega}{2} \left( p^2 + q^2 \right) + \omega_0\sqrt{1+\frac{\omega}{j\omega_0} \left(\frac{\gamma}{\gamma_c} \right)^2 q^2} \, m'.
\label{eq:bandas}
\end{equation}
These eigenvalues define effective semiclassical Hamiltonians for the slow bosonic variables, 
with  a number of important consequences:

{\em i)} In the region where the approximation is valid, the spectrum
of the Dicke model is divided in bands $E_{m'}(p,q)$, labelled by the quantum number $m'$. Each one has a semiclassical degree of freedom and hence is classically integrable. The ground-state band is that with $m'=-j$.

{\em ii)} The energy levels of the Dicke model are labelled by two quantum numbers, $E_{m',n}$. The first one, $m'$, identifies the band. The second one, $n$, determines the position of the energy level inside the band, which can be obtained following the Sommerfeld-Wilson-Ishiwara quantization rules
\begin{equation}
\oint_{C_{m'}} \, p \, dq = 2 \pi n,
\label{eq:req}
\end{equation}
where ${C_{m'}}$ is a closed curve for the effective Hamiltonian of the $m'$-esim band, and $n$ an integer number. The energy values $E_{m',n}$ for which the previous equality holds constitute a very good approximation to the energy levels of the Dicke model, as we shall see later.
 
{\em iii)} The expected value of any physical observable ${\mathcal
  O}(a^{\dagger},a; J_x, J_y, J_z)$, at a given energy $E$, can be
evaluated semiclassically
\begin{equation}
  \left< {\mathcal O} \right>=\frac{\int \, dp dq \, {\mathcal O} (p,q) \delta \left[ E - E_m(p,q) \right]}{\int \, dp dq \,\delta \left[ E - E_m(p,q) \right]},
\label{eq:semi}
\end{equation}
where ${\mathcal O} (p,q) = \left< j, m' \right| {\mathcal
  O}(a^{\dagger},a; J_x, J_y, J_z) \left| j,m' \right>$, and the bosonic operators $\left( a^{\dagger}, a \right)$ are written in terms of their classical limit $(p,q)$.

{\em iv)} $J_{z'}$ is an adiabatic invariant, semiclassical integral of motion. Thus, its quantum version
\begin{equation}
J_{z'} = \frac{J_z + \sqrt{\frac{\omega}{2 j \omega_0}} \frac{\gamma}{\gamma_c} \left( a + a^{\dagger} \right) J_x}{\sqrt{1+\frac{\omega}{2 j  \omega_0} \left( \frac{\gamma}{\gamma_c} \right)^2 \left( a + a^{\dagger} \right)^2}}  
\end{equation}
is expected to be a very approximated Hamiltonian commuting operator.

As long as the method of {\em freezing} the slow bosonic variables holds, the Dicke model can be  separated in a set of  semiclassical one-degree of freedom integrable models
. A simple bound for this approximation can be obtained selfconsistently,  taking into account that it will be valid only if the characteristic frequency of the bosonic mode is much less than the characteristic frequency of the pseudospin degree of freedom.
In the normal phase, these characteristic frequencies are approximately given by  the non-interacting ones, therefore the  condition of validity of the approximation is  given by $\omega/\omega_o<<1$, but in the superradiant phase this is not longer the case and a careful analysis is necessary.  To 
estimate these frequencies  in the  superradiant phase,
we  consider the ground-state band
$E_{m'=-j}(p,q) = \omega p^2/2 + V_{m'=-j} (q)$, where $V_{m'=-j}(q)$ is a conservative  potential.
 Expanding the latter around its minimum, we obtain
\begin{equation}
V_{m'=-j} (q) =V_{min}+ \frac{\omega}{2} \left[ 1 - \left(\frac{\gamma_c}{\gamma} \right)^4 \right] \left(q - q_{\text{min}} \right)^2 + \cdots.
\end{equation}
Therefore, with the definition $f \equiv \gamma/ \gamma_c$, the low energy frequency of the bosonic mode reads $\omega_B = \omega \sqrt{1-f^{-4}}$. For the atomic frequency, we evaluate 
the precession frequency 
in the same minimum as before, giving rise to $\omega_A= \omega_0 f^2$. Thus, the ratio $\omega_A/\omega_B= (\omega_0/\omega) f^4/\sqrt{f^4-1}$ and  the condition $\omega_B \ll \omega_A$ entails
\begin{equation}
\frac{\omega}{\omega_0} \ll \frac{f^4}{\sqrt{f^4-1}} \approx f^2.
\label{eq:condition}
\end{equation}
Hence, the approximation, and all the physical conclusions coming from it, are expected to be valid in the strong coupling regime $\gamma/\gamma_c=f\gg 1 $ (including the resonant $\omega/\omega_0=1$ case) or in the close to critical case $f\gtrsim 1$ for small enough $\omega/\omega_0$. It is worth to remark that the former is the less accessible regime for 
 numerical diagonalization. Furthermore, the larger the coupling constant $\gamma$ and/or the smaller the ratio $\omega/\omega_0$, the wider the region of energies for which this scenario is expected to work. This is specially interesting because it has been recently shown that the limit $\omega / \omega_0 \rightarrow 0$ is equivalent to the thermodynamic limit $j \rightarrow \infty$, giving rise to QPTs and ESQPTs in the Rabi model \cite{Richi:15}. The results we show in this Letter suggest that the same behavior is expected for the complete Dicke model, with any finite number of atoms. 


{\em Numerical tests.-} Besides the analytical argument, a stringent test is required to show that the regular energy region of the Dicke model is (very approximately) described by the Born-Oppenheimer approach
. To do so, we fully diagonalize the Dicke model for two highly representative cases, {\em a)} the resonant case $\omega=\omega_0=1$, for which the two harmonic oscillator approximation has been tested \cite{Emary:03}, and which is the object of the majority of the current research (see, for example, \cite{Baumann:10}); and {\em b)} the off-resonant case $\omega=1$, $\omega_0=5$. We set
$f=\gamma/\gamma_c=5$ in the first case  and $f=\gamma/\gamma_c=3$ in the second one. In both cases  the condition (\ref{eq:condition}) is clearly  fulfilled.

\begin{figure}
\begin{tabular}{c}
\vspace{-2cm} \\
  \hspace{-1cm}\includegraphics[width=1.15\linewidth,angle=0]{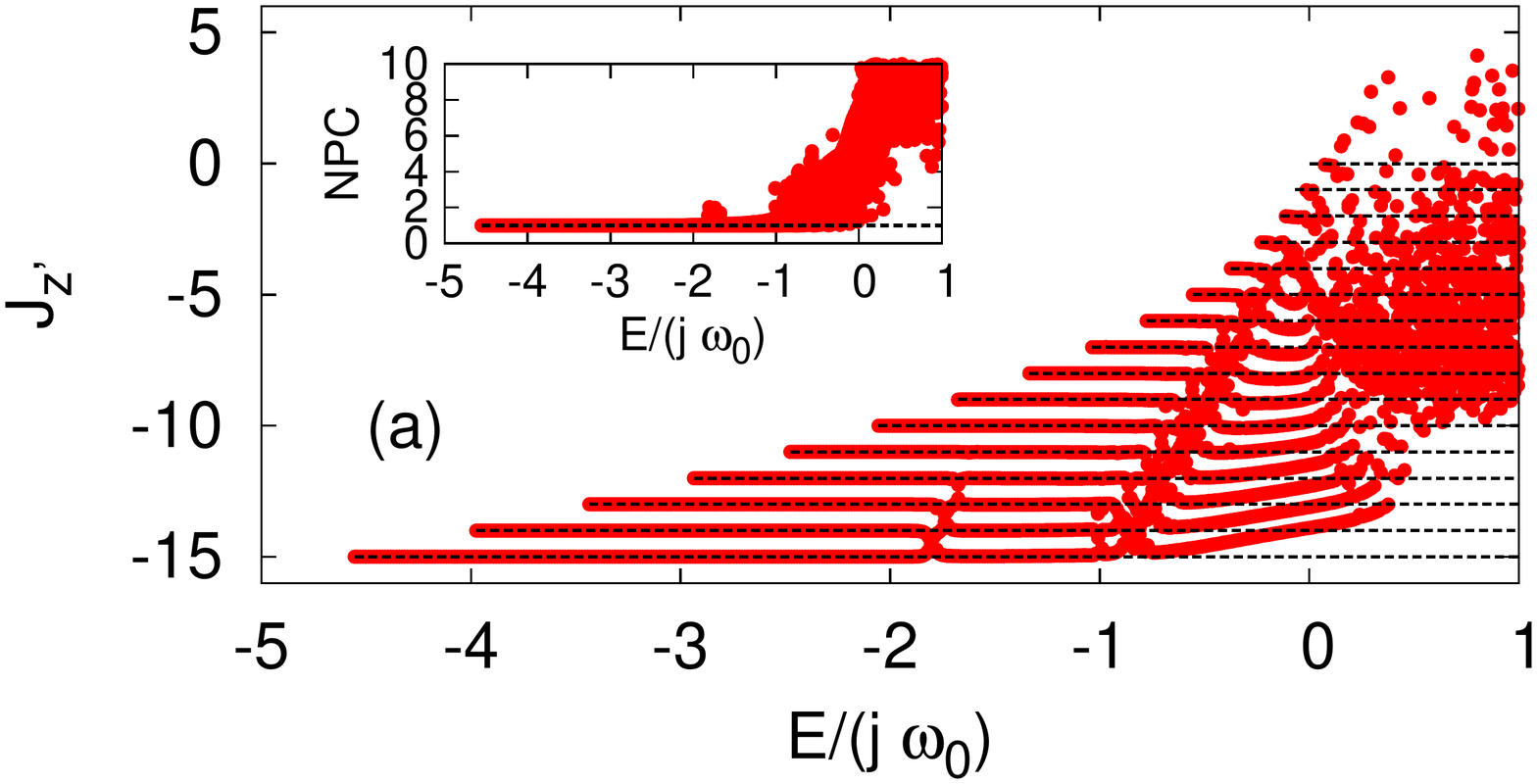} \\[-3.0cm]
 \hspace{-1cm} \includegraphics[width=1.15\linewidth,angle=0]{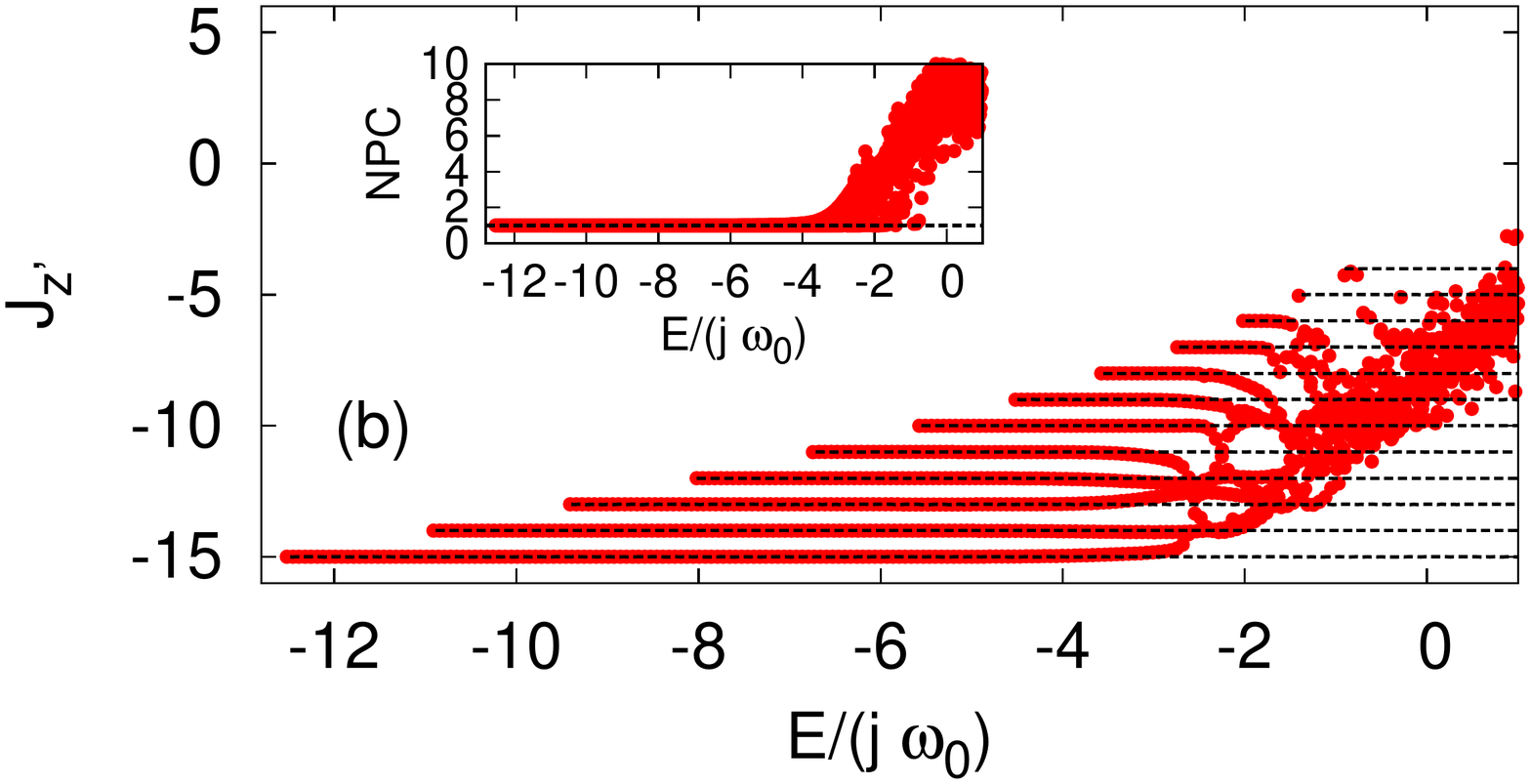} \\[-1.3cm]
\end{tabular}
\caption{Peres lattice for $J_{z'}$, with $j=15$.    Panel (a) shows the case $\gamma=3
  \gamma_c$, $\omega=1$, $\omega_0=5$; panel
  (b) $\gamma=5
  \gamma_c$ $\omega=\omega_0=1$. Horizontal lines show the eigenvalues $m'$
  of $J_{z'}$. Insets show the number of principal components (NPC) in
  both cases. The horizontal line represents $\text{NPC}=1$.}
\label{fig:jz}
\end{figure}

In Fig.\ref{fig:jz} we plot the Peres lattices \cite{Peres:84} for $J_{z'}$ with $j=15$
. They consist in drawing the expected value of representative observables in each eigenstate versus its respective eigenenergy, and have been used as a signature of quantum chaos \cite{Bastarrachea:14}. We can see in  both lattices, that 
in the low-energy region
, the expected value of $J_{z'}$ in
each eigenstate $|E_{i}\rangle$ is a very approximated integer number $m'$
, suggesting that $J_{z'}$ commutes with the corresponding projector $\left| E_i \right> \left< E_i \right|$, and thus $m'$ is a good quantum number. To prove this fact, we plot in the insets the number of principal components (NPC) of each Hamiltonian eigenstate, in the eigenbasis $\left| j,m'\right>$, $\text{NPC}=1/\sum_{m'} p_{m'}^2$, where $p_{m'}$ is the probability that an eigenstate lays in the eigenspace generated by $\left| j, m'\right>$. We can see that $\text{NPC} \approx 1$ in the low-energy region, showing that each eigenstate 
 within this region  belongs to the subspace generated by
$\left| j, m' \right>$, being $m'$ the corresponding quantum number. We can also see that this region extends until the ESQPT critical  energy $E/(j\omega_o)=-1$  for the case with $\omega_0=5$. In resonance, $\text{NPC} \approx 1$ up to $E/(j\omega_0) \sim -4$. 
The former results are fully compatible with the analysis presented in \cite{Bastarrachea:14, Chavez:16},
 where it is shown that in the case $\omega<\omega_0$, the onset of chaos occurs at energies above the ESQPT, whereas in the resonant case it occurs at energies below $E/(j\omega_o)=-1$. 
Furthermore, our results state that a necessary condition for the onset of chaos is that $J_{z'}$ ceases to be an integral of motion, and thus the eigenstates cannot be labelled by two quantum numbers anymore. These results provide a simple explanation for the approximated conservation rules observed in quenches leaving the system in the superradiant phase, very recently reported in \cite{Lobez:16}.

\begin{figure}
\begin{tabular}{c}
\vspace{-2.5cm} \\
 \hspace{-1cm} \includegraphics[width=1.15\linewidth,angle=0]{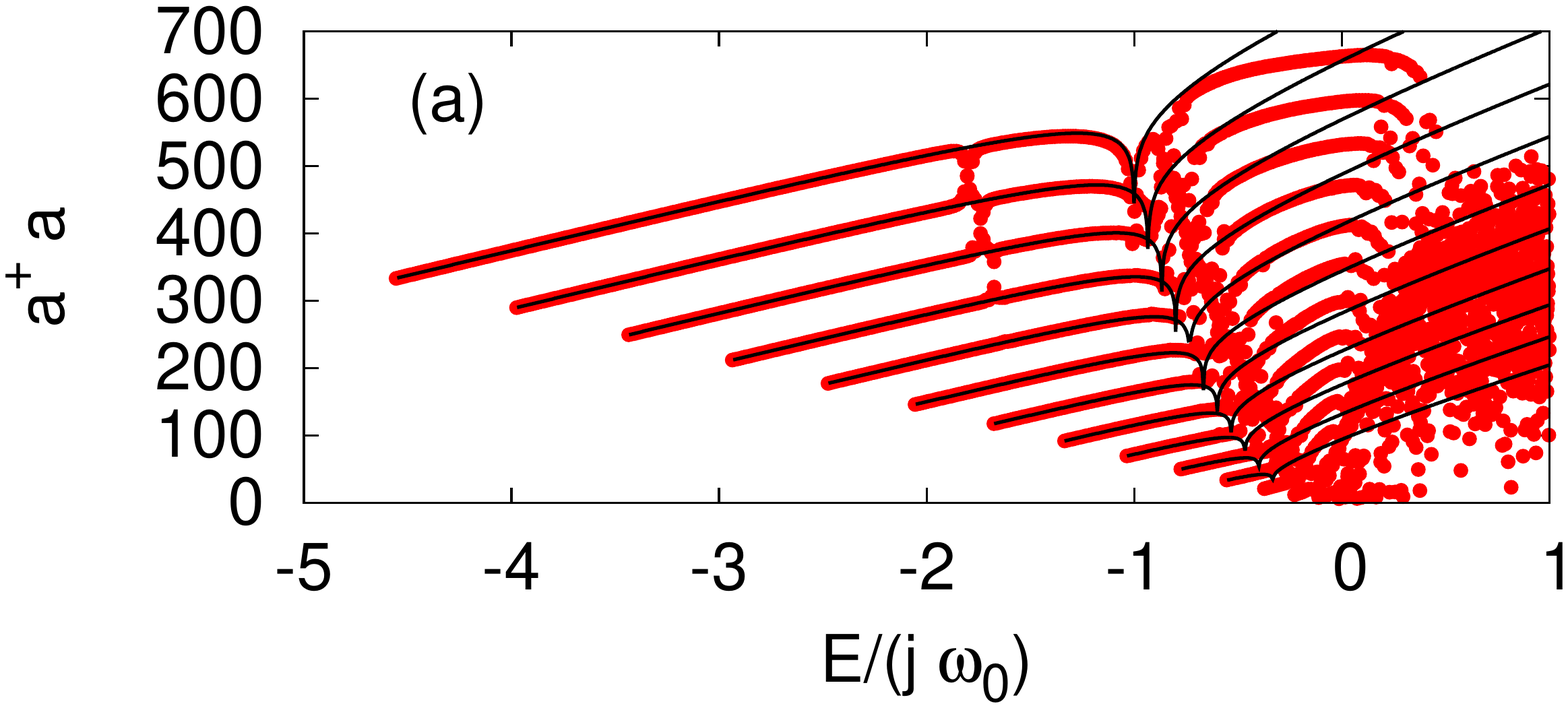} \\[-4cm]
  \hspace{-1cm}\includegraphics[width=1.15\linewidth,angle=0]{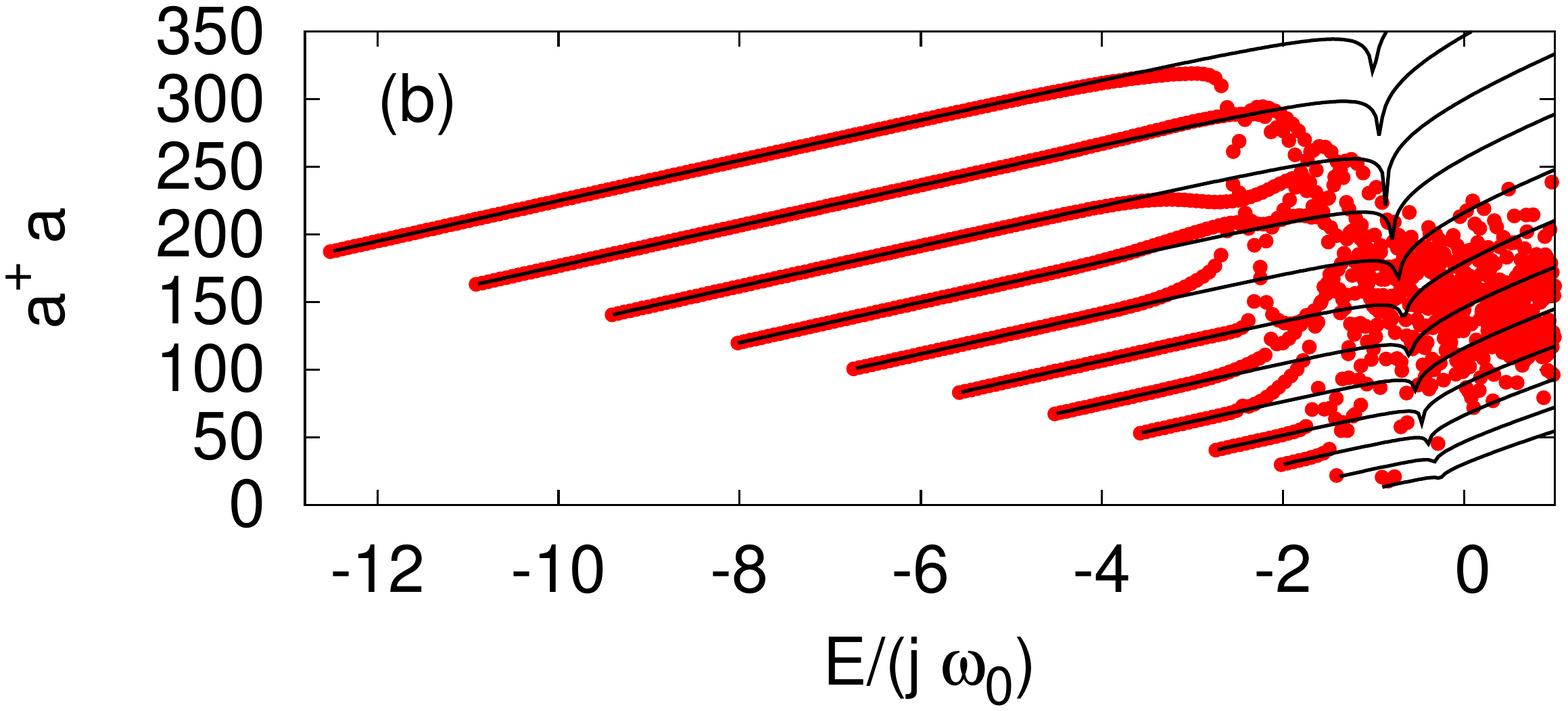} \\[-1.2 cm]
\end{tabular}
\caption{Peres lattice for $a^{\dagger} a$, with $j=15$. Circles (red online) show the exact results for the Dicke
  model. Solid lines represent the semiclassical description
  (\ref{eq:semi}). Panel (a) shows the case $\gamma=3
  \gamma_c$ with $\omega=1$, $\omega_0=5$;
  panel (b), $\gamma=5
  \gamma_c$ with $\omega=\omega_0=1$.}
\label{fig:na}
\end{figure}

In Fig.\ref{fig:na} we show the Peres lattices for the number of bosons $a^{\dagger} a$, together with the semiclassical description given by Eq. (\ref{eq:semi}). We can see that the region of applicability of Eq. (\ref{eq:semi}) coincides with the region in which $J_{z'}$ is a good integral of motion. This fact confirms our conclusion saying that the spectrum of the Dicke model is divided in independent bands, each one having one semiclassical degree of freedom, within the regular region. This is specially interesting for the case with $\omega_0=5$, shown in panel (a). In the upper band we can see a dip at $E/(j \omega_0) = -1$, which is properly reproduced by the semiclassical description; similar features are seen in the second and third bands, though in a less clear way. This is a neat signature of an one-degree of freedom ESQPT, characterized by logarithmic singularities in the density of states and expected values of representative observables \cite{Perez-Fernandez:11,Cejnar:07}; the same kind of ESQPT has been recently reported in the Rabi model \cite{Richi:15}. So, results shown in Fig. \ref{fig:na} suggest that every band has its own ESQPT, provided that the integrable region extends up to the corresponding energy value. This issue will be treated in detail elsewhere.

\begin{figure}
\begin{tabular}{c}
\vspace{-2.5cm} \\
 \hspace{-1cm}  \includegraphics[width=1.1\linewidth,angle=0]{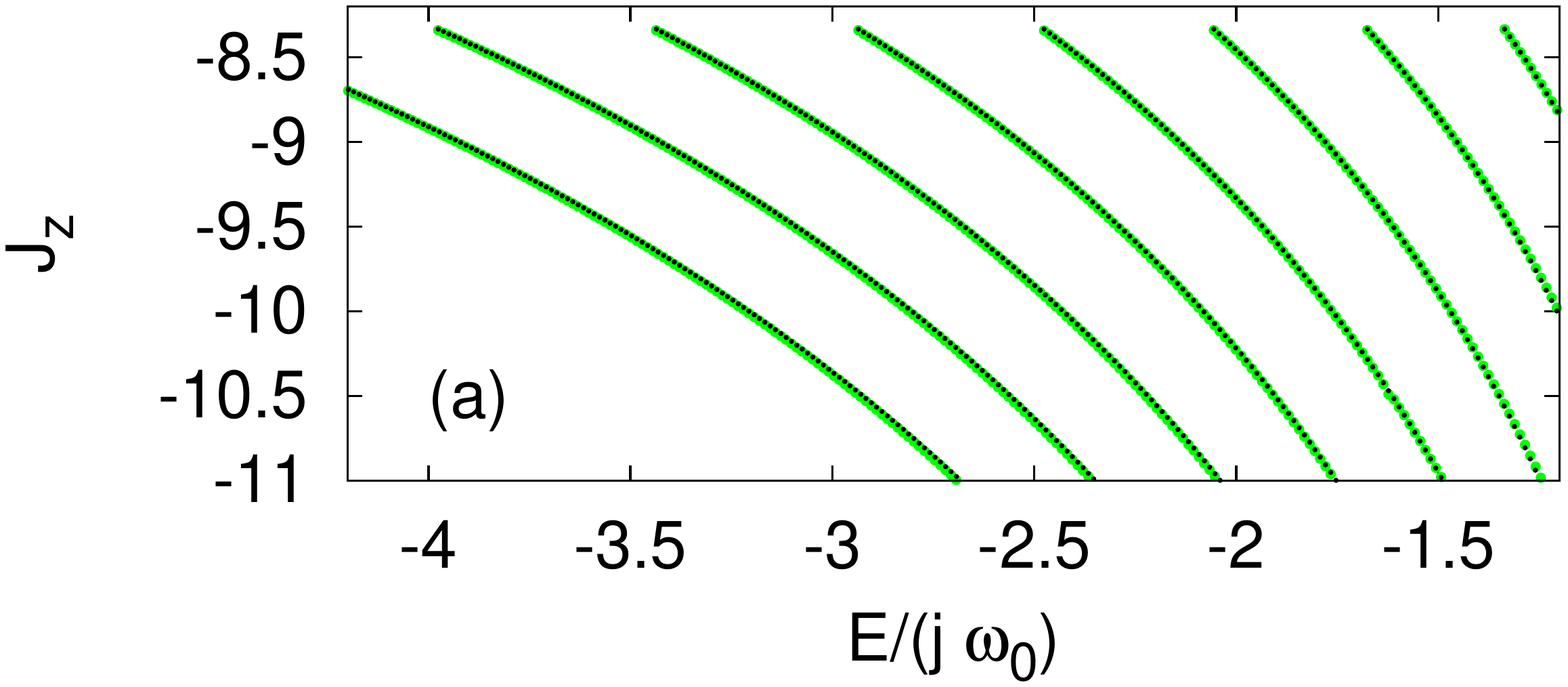} \\[-3.5cm]
  \hspace{-0.7cm} \includegraphics[width=1.05\linewidth,angle=0]{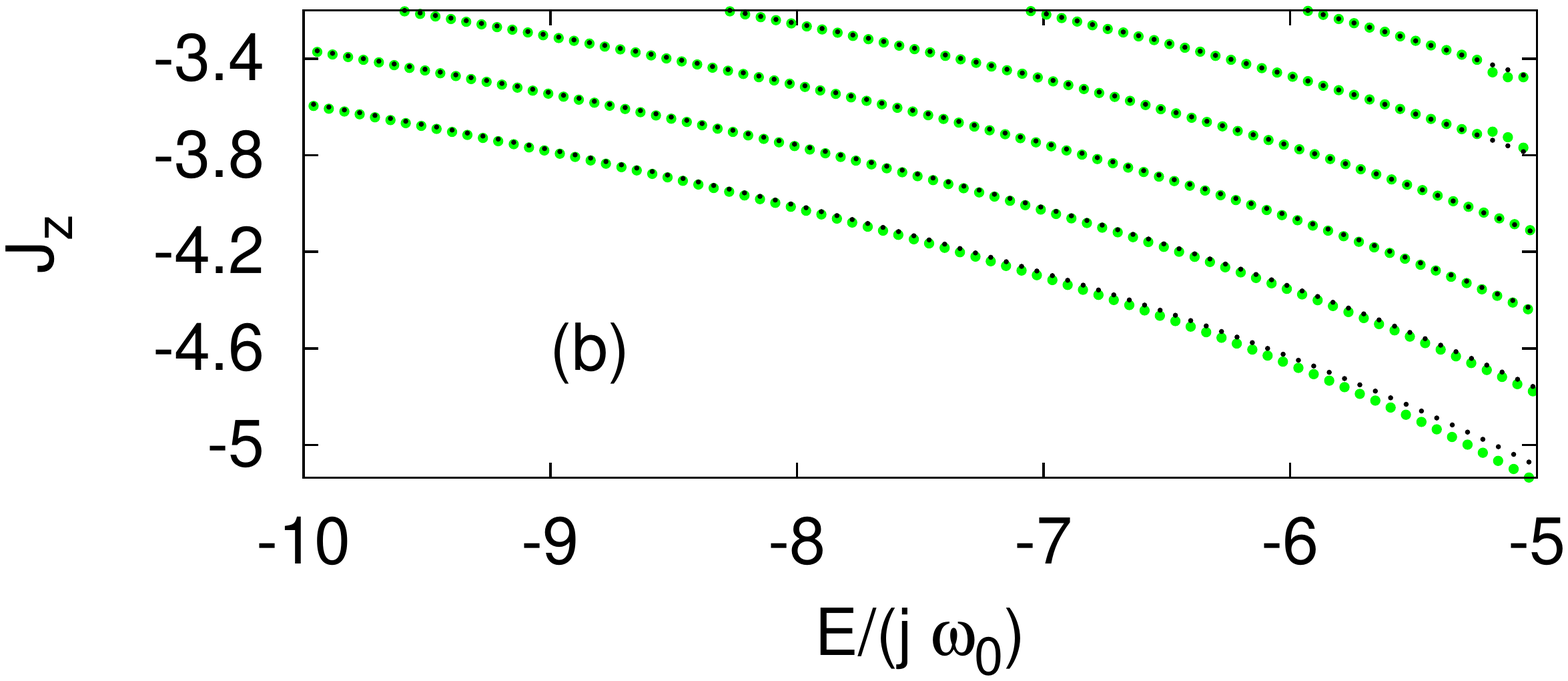} \\[-1.7cm]
  \hspace{-0.7cm} 
  \includegraphics[width=1.05\linewidth,angle=0]{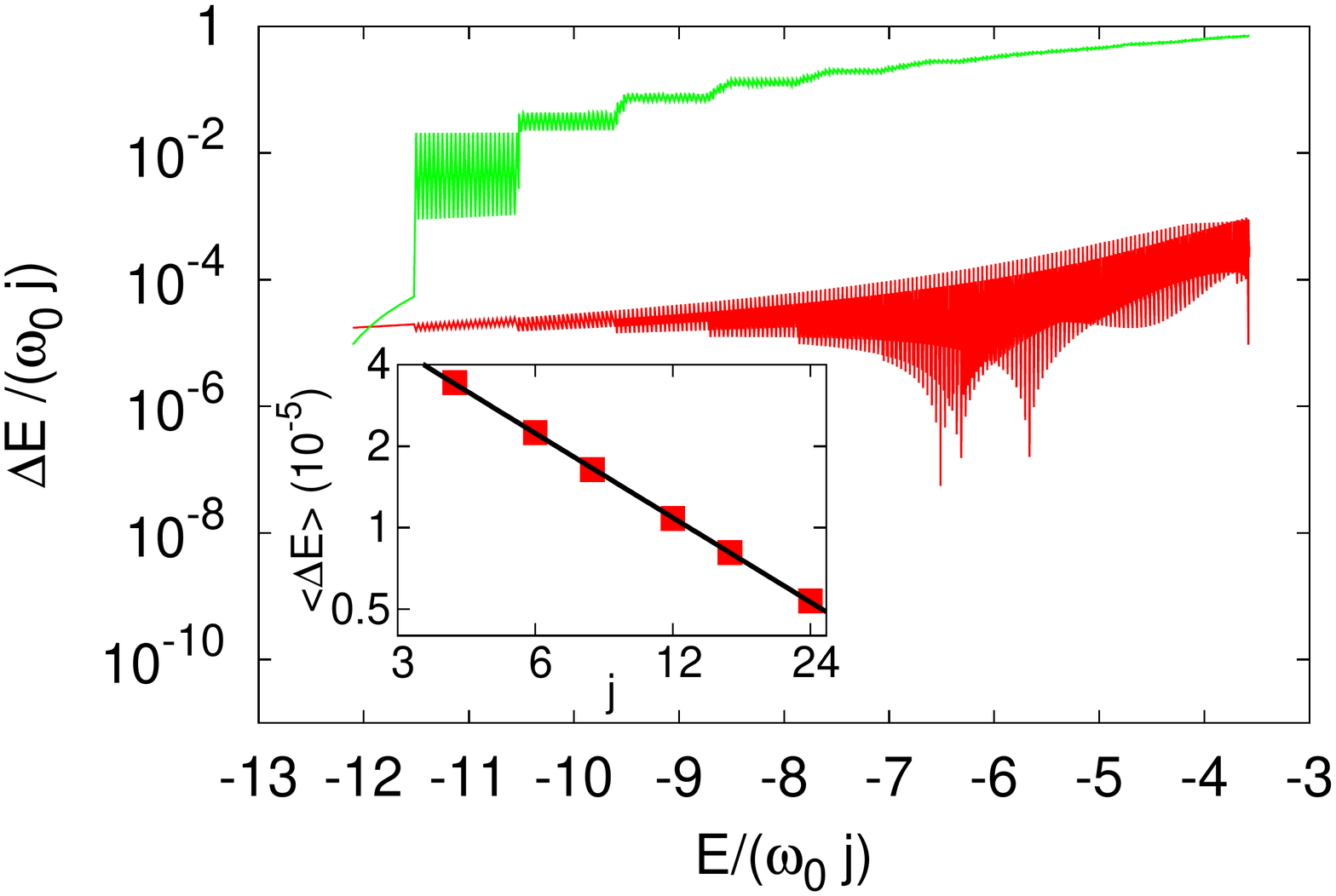}\\[-1.4cm]
  \hspace{0cm}
\end{tabular}
\caption{Comparison between the exact spectrum and the requantization
  procedure, Eq. (\ref{eq:req}). Panel (a) shows the Peres lattice for
  $J_z$ in the case $j=15$ with $\omega=1$, $\omega_0=5$ and $\gamma=3\gamma_c$. Light (green online) circles display the exact result; black circles, the result of Eq. (\ref{eq:req}). Panel (b) shows the same results for the case $j=15$ with $\gamma=5\gamma_c$ and $\omega=\omega_0=1$. Panel (c)  shows the  differences between the BOA requantization and exact spectrum as a function of energy (red online) in the resonant case. Similar results are shown for the harmonic approximation (green online). Inset shows  the average relative error for all energy levels within $-8 \leq E/(j \omega_0) \leq -6$  and system sizes between $j=5$ and
  $j=24$, in a log scale. 
  Solid black line represents a power-law fit
  $\langle\Delta E\rangle \propto j^{-\alpha}$, with $\alpha \approx 1.03$. }
\label{fig:comp}
\end{figure}

In Fig.\ref{fig:comp} we display the results of the requantization procedure, Eq. (\ref{eq:req}). In panels (a) and (b) we compare the exact Peres lattice for $J_z$ (not to be confused with the integral of
motion $J_{z'}$) depicted by light large circles (green online), with the values given by the semiclassical approximation, Eq.(\ref{eq:semi}), at the energies requantized by means of Eq.(\ref{eq:req}), depicted by  black small circles; we restrict ourselves to a smaller energy window than in Figs.\ref{fig:jz} and \ref{fig:na} to better appreciate the comparison. Requantization works almost perfectly for the case with
$\omega_0=5$. In the resonant case we can see slight differences in large energies,
 with the differences
laying in the vertical axis ($J_z$) and not in the horizontal one (eigenenergies). In panel (c)
we present the accuracy of the requantization procedure for the resonant case, as a function of energy in the main panel and as a function of the  the system size in the inset. In main panel the error of the requantized energy $\Delta E = \left| \left( E_{\text{req}} -E_{\text{exact}} \right) / E_{\text{exact}} \right|$ is plotted as a function of energy (red online).  As a reference, we plot  (green online) the same calculation for the two harmonic oscillator approximation \cite{Emary:03}. In the very low energy region both approximations give comparable accuracy, but the harmonic approximation breaks down rapidly as energy  increases ($E/(\omega_0 j)\sim -11$), contrary to the BOA results which gives an accurate description of the exact spectrum until very large excitation energies. The accuracy of the BOA as a function of the number of atoms is shown in the inset, we plot with squares (red online) the average error
of the requantized energy for different number of atoms. This average has been calculated including all the energy levels in  $E/(j \omega_0) \in \left(-8, -6 \right)$, a region in which $J_{z'}$ is a good integral of motion. Our main conclusion is that the error $\langle\Delta E\rangle$ clearly decreases with the system size. Indeed, we draw with a solid line the numerical fit to a power-law scaling $\left<\Delta E\right> \propto j^{-\alpha}$, with $\alpha = 1.03$. This result allows us to conjecture that $\Delta E \rightarrow 0$ as $j\rightarrow \infty$, but more work is needed to prove this statement. We have verified that, a similar exercise for the harmonic approximation, not only gives  much larger errors (more than two orders of magnitude) but also  they do not decrease with the system size. 


{\em Conclusions.-} By using a slow-fast motion, Born-Opppenheimer
approximation, we have been able to describe the regular (non-chaotic)
energy regime of the Dicke model 
in the superradiant phase, and to establish the region in parameter
space where the approximation is
valid. 
The very approximate second integral of motion, $J_{z'}$, was
identified, allowing to label with two quantum numbers, $E_{m',n}$,
all the energy levels within the regular energy region. By means of
stringent numerical tests, we have tested the accuracy of this
result. We have shown that the corresponding semiclassical picture
provides a very good estimate for the expected values of
representative physical observables. And the same for the energy
levels $E_{m',n}$, obtained with a requantization method which do not
require the numerical diagonalization of the Hamiltonian. Since the ratio of effective  atom to  boson  frequencies 
($\omega_A/\omega_B$), increases as a function of $f=\gamma/\gamma_c$
and $\omega_0/\omega$, making the BOA a better approximation, it is
expected that the regular energy interval of the Dicke model increases
accordingly to these same
variables.  
The  relevant resonant case is included in the region of
applicability of the BOA in the superradiant phase, contrary to the naive expectation that is
only valid if $\omega_o/\omega\gg 1$.  It has not only been shown that
the BOA applies at energies far above the region in which the double
harmonic oscillator approximation holds \cite{Emary:03}, even reaching
the critical energy of the ESQPT when the Dicke model is out of
resonance, but also that the BOA's accuracy increases with the number
of atoms. We think these results gives new insights into the wide
regular region of the Dicke model, by extending the theoretical tools
to describe what the quadratic  approximation, obtained in turn from 
a Holstein-Primakoff one, leaves unexplained. What is more, the
knowledge of this second integral of motion is necessary to study
properly the model, and can be very useful for the research of
non-equilibrium dynamics and the route to chaos, both theoretically
and experimentally.

\begin{acknowledgements} 
The authors thank J. Dukelsky for enlighting discussions, and J. Larson for pointing out very relevant references. A. R. is supported by Spanish Grants No. FIS2012-35316 and FIS2015-63770-P (MINECO/FEDER), M. A. B-M and S.L-H acknowledge financial support from mexican CONACyT project number CB166302. S.L-H. acknowledges financial support from CONACyT  fellowship program for sabbatical leaves. 
\end{acknowledgements}

\end{document}